\documentclass[12pt]{article}
\let\section=\subsection     \let\subsection=\subsubsection                
\usepackage{graphicx}
\usepackage{amsmath}
\def\bom{\boldsymbol\omega}

\begin{document}
\begin{center}
   {\large \bf In-medium properties of the $\bom$ meson}\\[2mm]
   {\large \bf through $\bom$ photoproduction in nuclei}\\[5mm]
   P.~M\"uhlich, T.~Falter and U.~Mosel \\[5mm]
   {\small \it  Institut f\"ur Theoretische Physik \\
   Universit\"at Giessen \\
   Heinrich-Buff-Ring 16, D-35392 Giessen, Germany \\[8mm] }
\end{center}

\begin{abstract}\noindent
   The feasibility to study the in-medium properties of the $\omega$ meson 
   in photon induced reactions on nuclei is investigated by means of a semiclassical
   transport model. The calculations are dedicated to an experiment presently being 
   analyzed by the TAPS/Crystal Barrel collaborations. A full coupled-channel treatment of 
   final state interactions allows a quantitative analysis of the influence of absorption and 
   elastic scattering of the produced hadrons and likewise the application of kinematic cuts on
   the observables. Within the scope of our calculations, the preliminary data by the TAPS/Crystal Barrel
   collaborations indeed seem to indicate a modification of the isoscalar spectral density in the strongly
   interacting environment.
\end{abstract}

\section{Introduction}
The possibility of a direct locking of the masses of the light vector mesons and the chiral quark condensate \cite{Brown:1991} has opened a unique method to determine experimentally to what degree chiral symmetry is restored in the strongly interacting medium. Whereas the universal scaling law proposed by Brown and Rho \cite{Brown:1991} is not expected to hold in a strict sense, also QCD sum-rules have brought forward the idea of dropping vector meson masses as density and temperature increase \cite{Hatsuda}. However, a more recent analysis of QCD sum rules has shown that in such a framework no rigorous proof of decreasing masses can be given and that the sum rules could also be fulfilled by keeping the masses fixed and increasing the width of the hadrons in medium \cite{Leupold:1998QCDSR}. From a completely different point of view sizable changes of the vector meson in-medium properties have also been predicted in purely hadronic scenarios \cite{Klingl:1997,Post:2003hu} without being directly linked to chiral symmetry restoration.

The present work is dedicated to an experiment presently being analyzed by the TAPS/Crystal Barrel collaborations at ELSA \cite{TAPS}. The considered reaction is inclusive photoproduction of $\omega$ mesons in nuclei which are detected via their semi-hadronic decay to $\pi^0\gamma$ pairs. The evident advantages of this reaction channel are the following. Due to its electromagnetic coupling to the nucleon, the reaction probability of the photon is almost the same for all nucleons of the target nucleus. Hence, all densities of this static density distribution are probed. The $\omega$ meson itself forms a unique probe since on the one hand it is expected to experience sizable in-medium changes and on the other hand should even at nuclear saturation density exhibit a proper resonance structure to which a quasiparticle mass and width can be addressed. The advantage of detecting the $\pi^0\gamma$ decay mode is the large branching of the $\omega$ to this channel ($8.7\cdot10^{-2}$) and the fact that there is only one strongly interacting particle in the final state, giving rise to less pronounced final state interactions as compared to purely hadronic decay channels \cite{Muhlich:2003tj}.

In the following section we describe very briefly the transport model applied. In Section \ref{results} we focus on the reduction of rescattering background and the visibility of in-medium changes of the $\omega$ meson. Finally, we draw our conclusions in Section \ref{conclusions} together with a short outlook on the present experimental situation.

\section{Model}
Applying the impulse and the local density approximation, the mass differential cross section for photoproduction of $\pi^0\gamma$ pairs off nuclei can be written as follows:
\begin{eqnarray}
\left(\frac{d\sigma}{d\mu}\right)\sim
\int^{NV} d^3r\int^{{p_F({\bf r})}}d^3p_N\int dm~\delta(\mu-M_{\pi^0\gamma}){\mathcal{N}_{\pi^0\gamma}(M_{\pi^0\gamma},...)}\nonumber\\
\times~\sigma_{\gamma N\to\omega X}(m)a_{\omega}({\bf r}, {\bf p}, m)\frac{\Gamma_{\omega\to\pi^0\gamma}(m)}{\Gamma_{\omega}^{\mathrm{tot}}},
\end{eqnarray}
where $\sigma_{\gamma N\to\omega X}$ is the total cross section for $\omega$ photoproduction of an individual nucleon and $a_{\omega}$ is the $\omega$ spectral function. The multiplicity factor $\mathcal{N}_{\pi^0\gamma}$ specifies the asymptotic number of $\pi^0\gamma$ pairs with invariant mass $M_{\pi^0\gamma}$. Whereas without final state interactions this factor would consist of a $\delta$-function, in our calculations $\mathcal{N}_{\pi^0\gamma}$ is determined by means of the coupled-channel transport model, accounting for particle propagation, elastic and inelastic scattering, absorption and decay of all involved particle species. The implementation of the transport model to photonuclear reactions is described in detail in Refs. \cite{Muhlich:2003tj,Muhlich:2002tu,Effenberger:1999ay}. For the elementary production process $\gamma N\to\omega X$ we include the exclusive and inclusive production mechanisms. Cross sections for inclusive production channels at higher energies which have not been measured so far are calculated by means of the hadronic event generator FRITIOF, to which the photon is coupled via the simple vector meson dominance model. This method also is described in Ref. \cite{Muhlich:2002tu}.

Within the BUU approach the propagation of medium modified particles is done by means of a phenomenological off-shell potential \cite{Effenberger:1999ay}. This potential shifts the mass of any particle from its in-medium value to a mass according to its vacuum spectral function as the local density vanishes. The medium modification of the $\omega$ meson is introduced by means of the low-density theorem, which yields a collisional broadening of about 40 MeV at nuclear saturation density and vanishing momentum. For the effective in-medium mass we use as a first approximation
\begin{eqnarray}
m^*=m_0+U_S({\bf r}),\quad U_S({\bf r})=-0.16 m_0\frac{\rho({\bf r})}{\rho_0}
\label{mass_shift}
\end{eqnarray}
as indicated e.~g. by the calculation of Ref. \cite{Klingl:1997}.

\section{Inclusive $\bom$ production in nuclei}\label{results}

\begin{figure}[b!]
\begin{center}
   \includegraphics[scale=.6]{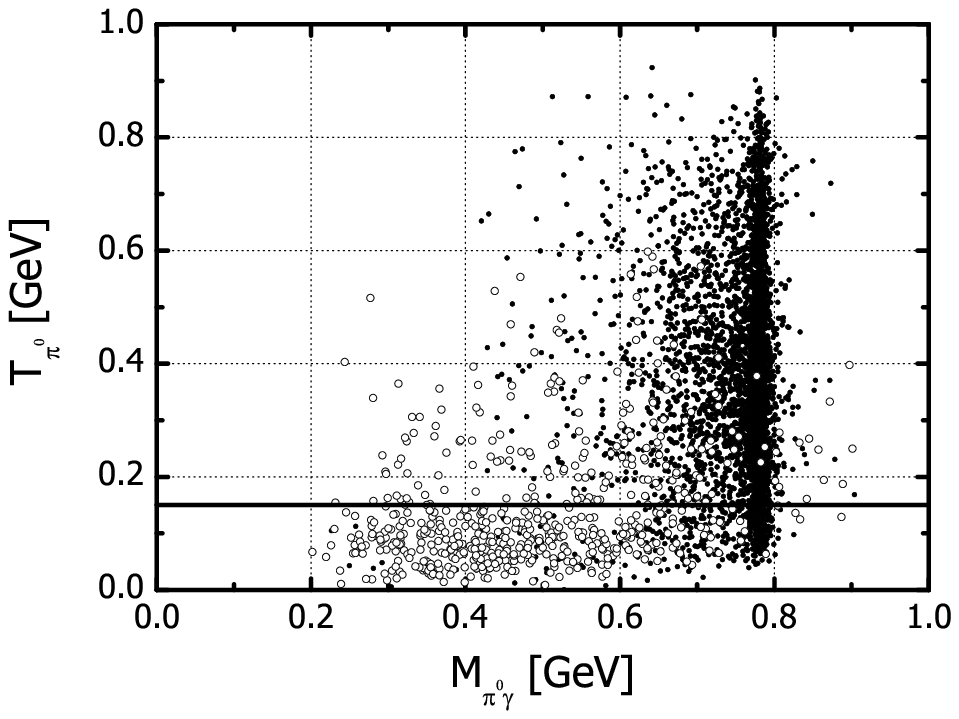}\\
   \parbox{\textwidth}
        {\footnotesize {\centerline\\
        Fig.~1: Pion kinetic energy versus $\pi^0\gamma$ invariant mass. 
        The horizontal line indicates the kinetic energy cutoff.}}
\end{center}
\vspace*{-.6cm}
\end{figure}

The most critical argument against measuring the $\omega$ meson through the $\pi^0\gamma$ decay channel is the strong final state interaction of the $\pi^0$. In Fig.~1 we show the $\pi^0$ kinetic energy versus the $\pi^0\gamma$ invariant mass for events where the $\pi^0$ propagates out of the nucleus without further interactions (solid circles) and events where the $\pi^0$ scattered from some nucleon in the surrounding medium (open circles). Due to the dominant excitation and decay of $\Delta$-resonances in intermediate energy $\pi N$ scattering, the kinetic energies of the scattered pions are accumulated around values of 100 MeV. This allows to reduce the rescattering background considerably by applying a cutoff on the $\pi^0$ kinetic energy as indicated by the horizontal line (Fig.~1). This is also described in Ref. \cite{Messchendorp:2001pa}.

\begin{figure}
\begin{center}
   \includegraphics[scale=.8]{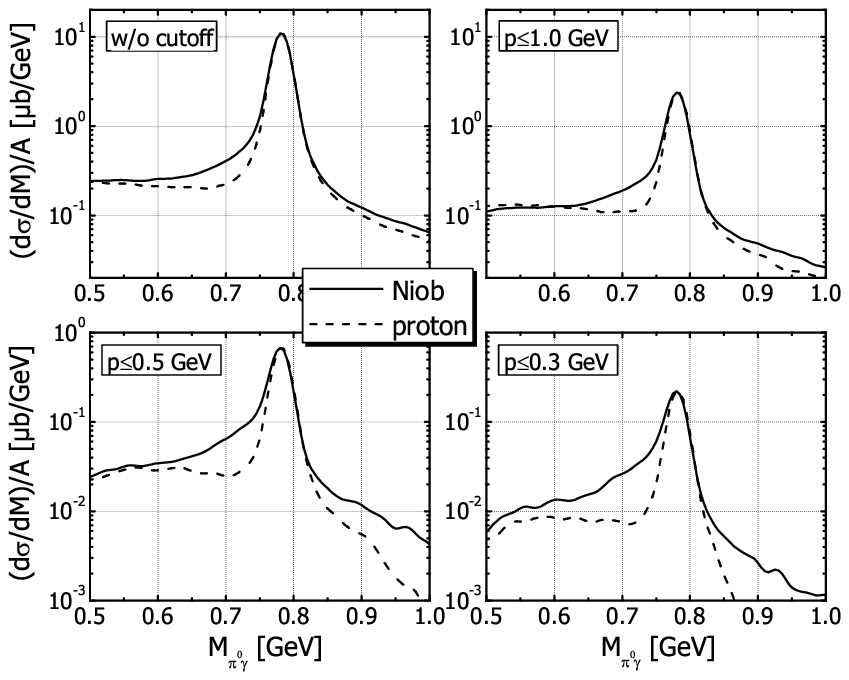}\\
   \parbox{\textwidth}
        {\footnotesize {\centerline\\
        Fig.~2: Comparison of invariant mass distributions obtained from proton and Niobium targets for several values of the three momentum cutoff. The cross sections on the proton are normalized to the cross section on Niobium at the $\omega$ pole mass (782 MeV).}}
\end{center}
\vspace*{-.6cm}
\end{figure}

\begin{figure}[b!]
\begin{center}
   \includegraphics[scale=.8]{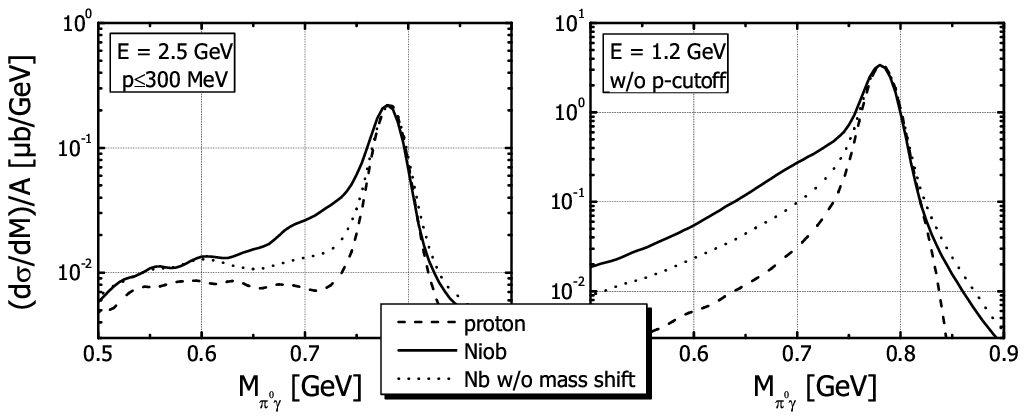}\\
   \parbox{\textwidth}
        {\footnotesize {\centerline\\
        Fig.~3: Results for the invariant mass distributions on proton and Niobium targets for 1.2 GeV and 2.5 GeV photon energy. The dotted curve contains collisional broadening only.}}
\end{center}
\vspace*{-.6cm}
\end{figure}

Minimizing the decay length of the produced $\omega$ mesons enhances the a\-mount of 
in-medium decays. This implies, that the momentum of the produced $\omega$ mesons should be as low as possible. Whereas the average $\omega$ momentum is smallest for very low photon energies, low $\omega$ momenta are also populated for higher beam energies, e.~g. 2.5 GeV which is the maximum energy used in the TAPS/CB experiment \cite{TAPS}. In this case a three momentum cutoff has to be applied to the detected $\omega$ mesons in order to get rid of the high momentum part of the spectrum, which carrys no information about the $\omega$ in-medium mass. The effect of such a cutoff is demonstrated in Fig.~2, where the $\pi^0\gamma$ mass spectrum on proton and Niobium targets is compared for various cutoff values. With a decreasing cutoff we find more pronounced deviations from the vacuum situation due to the higher average densities probed.

In Fig.~3 we show the results for 1.2 GeV and 2.5 GeV photon energy. The figure contains two additional curves, which are calculated without the shift of the $\omega$ mass, i.~e. containing collisional broadening only. All curves are normalized to the one containing the dropping $\omega$ mass. These results show, that most of the signal stems from the downward shift of the $\omega$ mass and, therefore, reveals a non-trivial in-medium change of the $\omega$ spectral density in nuclei.

\begin{figure}
\begin{center}
\parbox{8cm}{
   \includegraphics[scale=.6]{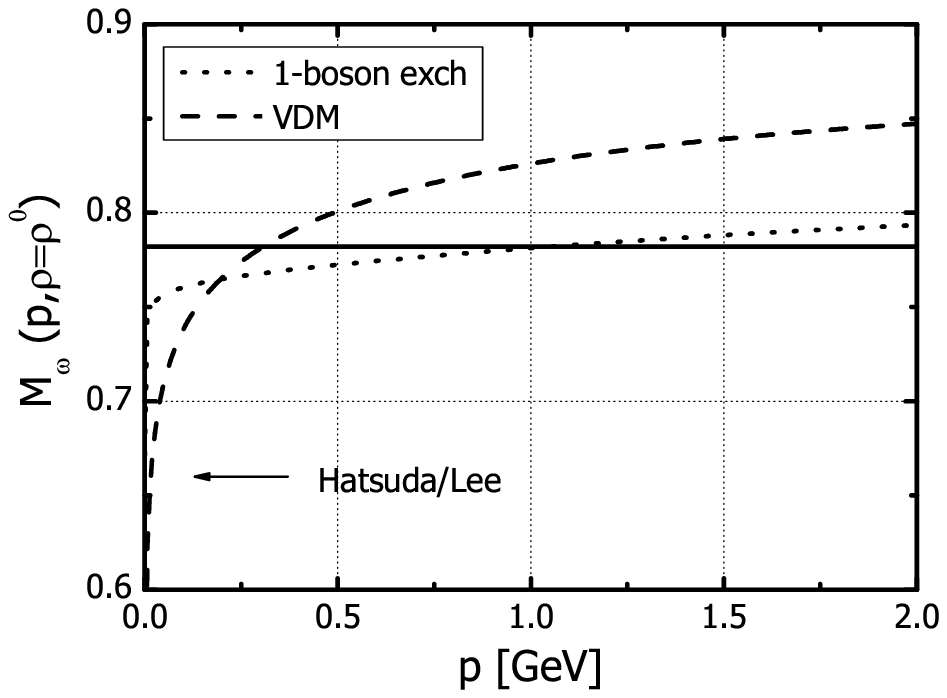}
   }\parbox{6cm}{
   \includegraphics[scale=.6]{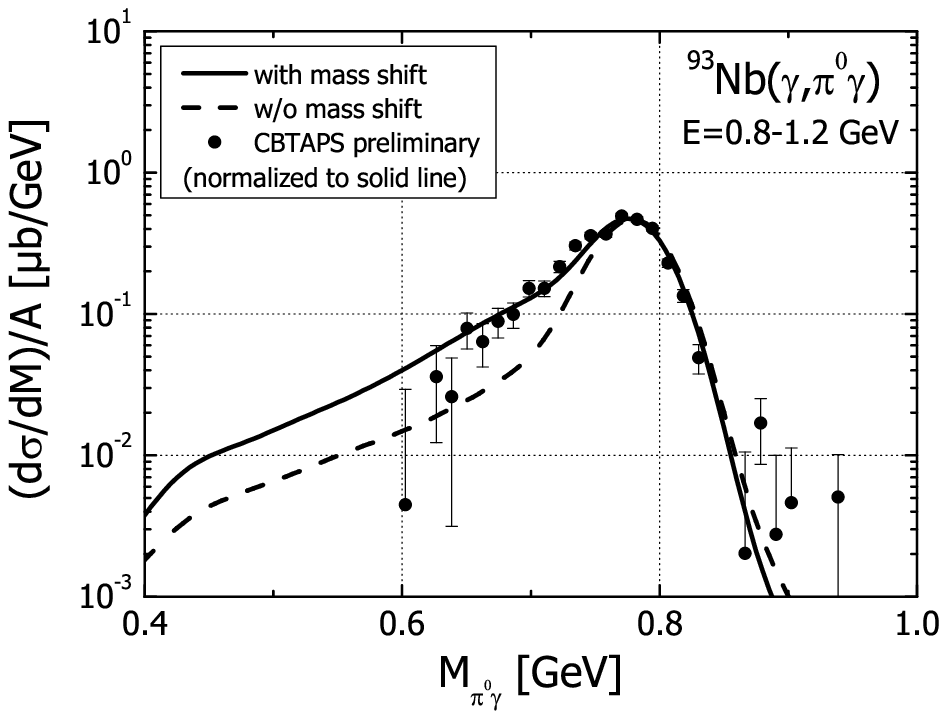}
   }\\
   \parbox{6cm}
        {\footnotesize {\centerline\\
        Fig.~4: Effective $\omega$ mass as function of the $\omega$ three momentum at nuclear saturation density.}}\hfill\parbox{6cm}
        {\footnotesize {\centerline\\
        Fig.~5: Invariant mass spectrum on a Niobium target with preliminary experimental data from \cite{TAPS}.}}
\end{center}
\vspace*{-.6cm}
\end{figure}

The results shown so far have been obtained with the purely density dependent mass shift of Eq.~\ref{mass_shift} which in principle should be valid only for $\omega$ mesons at rest. However, real and imaginary part of the $\omega$ self-energy in matter are connected by a dispersion relation, which leads within the low-density expansion to a momentum dependent mass shift that is driven by the energy dependence of the $\omega N$ forward scattering amplitude:
\begin{eqnarray}
\label{realpart}
{m^*}^2=m^2-4\pi\mathcal{R}e f_{\omega N}(E,\theta=0)\rho_N
\end{eqnarray}
This mass shift has been estimated employing two different assumptions on the $\omega N$ total cross section, one obtained by the vector meson dominance relation to $\omega$ photoproduction and the other one taken from Ref. \cite{Lykasov:1998ma}. The results shown in Fig.~4 exhibit a qualitatively similar behavior: a considerably reduced mass at zero momentum which very rapidly increases to a slightly repulsive behavior for large $\omega$ momenta. As one expects the results on the $\pi^0\gamma$ mass distribution obtained with this momentum dependence are very similar to the results without the $\omega$ mass shift, depicted by the dotted curve in Fig.~3. Such a momentum dependence could only be studied by considerably decreasing the cutoff on the $\omega$ three momentum well below values where the $\omega$ mass shift becomes small.

\section{Conclusions}\label{conclusions}

In Fig.~5 we show our results for the averaged photon energy bin $E_{\gamma}=(0.8-1.2)$ GeV obtained without the dispersive momentum dependence, i.~e. with the mass shift according to Eq.~\ref{mass_shift}, together with the preliminary experimental data from \cite{TAPS}. Disregarding the preliminary nature of the experimental data, from this comparison one would conclude that there is a sizable modification of the $\omega$ spectral density in nuclei with a shift of spectral strength to lower invariant masses (at least for $\omega$ momenta around $(200-400)$ MeV, to which such a measurement is sensitive). At the same time the question arises, why the dispersive analysis of the real part of the $\omega$ self energy cannot reproduce the experimental data, maybe either because of the more complex structure of the $\omega N$ scattering amplitude or because of the non-validity of the applied low-density approximation (see, e.~g.~\cite{Post:2003hu}). On the other hand, more data with lower momentum cuts might help to fix the momentum dependence of the $\omega$ self-energy. A better understanding of the experimental background is mandatory in order to unambiguously confirm the mass shift. More work, both experimental and theoretical, is required to clarify this issue.

The authors acknowledge valuable discussions with D.~Trnka and V.~Metag.
This work has been supported by DFG.

\end{document}